\documentclass[12pt]{article}
\usepackage[a4paper]{geometry}
\usepackage{anysize}
\marginsize{2cm}{2cm}{2cm}{2cm}
\usepackage[english]{babel}
\usepackage[T1]{fontenc}
\usepackage[utf8]{inputenc}
\usepackage{amsmath,amsthm,amssymb,xfrac}
\usepackage{graphicx}
\usepackage{physics}
\usepackage{array}
\usepackage{dsfont}
\usepackage{hyperref}
\usepackage{todonotes}
\usepackage{pdfpages}
\usepackage{mathtools}
\usepackage[deletedmarkup=xout]{changes}
\usepackage{authblk}
\usepackage[backend=biber, style=numeric-comp, sorting=none]{biblatex}
\usepackage{csquotes}
\usetikzlibrary{arrows.meta,positioning,decorations.pathmorphing}
\definechangesauthor[color=blue]{Barna}

\definechangesauthor[color=red]{Helmut}

\title{Positronium Laser Deceleration, Cooling and Trapping}
\author[1]{Barna Mendei\footnote{\href{mailto:Barna.Mendei@uibk.ac.at}{Barna.Mendei@uibk.ac.at}}}
\author[1]{Helmut Ritsch}
\date{\today}
\affil[1]{Department of Theoretical Physics, University of Innsbruck}

\begin{document}

\maketitle

\begin{abstract}

We numerically study laser deceleration, cooling and trapping of host fast ortho-positronium (Ps) bunches as produced in silica microchannel targets using positron bunches from a Surko trap. While Doppler laser cooling routinely allows to cool neutral atom species below milli-Kelvin kinetic temperatures, the fast ground state annihilation and the high recoil shift to line-width ratio in Ps pose new challenging limitations on timing and  the required laser power. Earlier theoretical work and first experiments already exhibit viable prospects for radiation pressure cooling to create unprecedented cold and dense Ps ensembles. Our extended numerical studies confirm very good prospects to simultaneously decelerate, cool and optically trap Ps atoms in a standard Doppler cooling geometry on the $1^3\mathrm{S}$--$2^3\mathrm{P}$ transition. Including the full Zeeman state manifolds and adding additional transverse lasers on the $2^3\mathrm{P}$--$3^3\mathrm{D}$ transition improves timing and final temperature to allow for optical trapping at the end. The required laser powers and geometry parameters to implement effective slowing and trapping within a time frame, where the majority of atoms is not annihilated, is in reach of current technology of $\mathrm{mJ}$ energy pulses of approximately $100\ \mathrm{ns}$ duration. Future extensions to collectively enhance cooling and trapping in optical cavities or hollow-core fibres should finally allow fast preparation of ultra-cold Ps systems as future basis of superradiant lasing and Bose--Einstein condensates.

\end{abstract}

\newpage

\section{Introduction}

Positronium (Ps), the bound state of an electron and a positron, has long been a subject of fundamental interest due to its unique properties as a purely leptonic system containing anti-matter as a pure QED system~\cite{bass2023colloquium}. This gives great potential for precision spectroscopic QED tests and beyond in dark matter detection, gamma ray amplification and even for the realization of Bose--Einstein condensation (BEC) of a purely leptonic system~\cite{liang1988laser}. To implement these fascinating options, positronium laser cooling has been theoretically proposed already several decades ago~\cite{liang1988laser, iijima2000laser} with more detailed studies lately~\cite{zimmer2021positronium}. The theoretical suggestions were soon followed by early experimental efforts~\cite{kumita2002study, hirose2002laser} as a clear proof of principle. For precision tests of Ps energies two photon spectroscopy of the Ps $1\mathrm{S}$--$2\mathrm{S}$ line was already performed with surprising accuracy~\cite{fee1993measurement, cooke2015observation} complemented by measurements of the Ps Doppler profile in microtraps~\cite{cassidy2010positronium}.

Here we first summarize the history and state-of-the-art of Ps laser cooling. The first theoretical proposals for laser cooling of Ps emerged in the late 1980s, highlighting the feasibility of Doppler cooling using a $243\ \mathrm{nm}$ laser strongly red detuned to the $1\mathrm{S}$--$2\mathrm{P}$ transition. Ps Doppler cooling already targets the recoil temperature of $T_\text{rec}\approx148\ \mathrm{mK}$, which due to the small mass is higher than the standard Doppler limit $T_D\approx1.2\ \mathrm{mK}$. Such narrow line cooling, studied also for atomic clock transitions, can reach even below the recoil temperature~\cite{loftus2004narrow}.

First experimental efforts on Ps cooling in the 2000s focused on ortho-positronium (o-Ps) and demonstrated the cooling of Ps atoms using pulsed lasers and bunched positron beams~\cite{iijima2000laser, kumita2002study}. These studies established the feasibility of Ps cooling and identified some key challenges, such as the short lifetime of Ps ($142\ \mathrm{ns}$ in vacuum) and the need for very broad-linewidth lasers to mitigate Doppler broadening. Subsequent theoretical and experimental work explored the use of magnetic fields to enhance cooling efficiency and mitigate annihilation rates, with particular success at low and high field strengths~\cite{zimmer2021positronium}.

In recent years, significant progress has been made towards achieving efficient and scalable Ps cooling. Experimental demonstrations have shown one-dimensional Doppler cooling of Ps using broadband, chirped laser pulses, reducing the velocity distribution to $\approx1\ \mathrm{K}$ in $100\ \mathrm{ns}$~\cite{shu2024cooling}. Further advancements include the use of broadband alexandrite lasers at $243\ \mathrm{nm}$, which achieved substantial temperature reductions (from $380\ \mathrm{K}$ to $170\ \mathrm{K}$) in tens of nanoseconds, and the first broadband laser cooling of Ps within the AEgIS experiment at CERN~\cite{gloggler2024positronium, caravita2025laser}. These results represent a critical step towards the realization of Ps BEC and the development of antimatter-based quantum technologies. 

With the progress in new higher flux and denser Ps sources as e.g., realized via laser written Silica micro cavities, new and more detailed studies of slowing, trapping and cooling of Ps by pure optical means are needed to judge prospects for precision spectroscopy and many-body physics with a cold leptonic ensemble. The specific objective of this study was to explore the possibilities and experimental feasibility of Ps laser cooling. This involves the numerical analysis of different laser schemes and parameters.

Our work is organized as follows: we first introduce the theoretical models and methods used below in section~\ref{methods}. this includes the most important physical properties of Ps and an introduction to the semi-classical approach for our simulations based on two generic laser configurations. Section~\ref{results} presents initial conditions of our simulations and our main findings focusing on three issues: optimising on reaching the lowest overall temperature, highest population of non-annihilated atoms at the end, and at the time when temperature reached $1\ \mathrm{K}$. Finally in section~\ref{discussion} we discuss these results in the context of the practical experimental feasibility.

\section{Model and Methods}
\label{methods}

\subsection{The Positronium (Ps) atom}

Let us first review some basic properties of the Ps atom. It is the bound state of an electron and its antiparticle, the positron and thus is composed of $50\%$ antimatter. Its  quantum mechanical description is analogous to the hydrogen atom, but it is a purely leptonic bound system without a nucleus composed solely of leptonic point particles~\cite{stroscio1975positronium}. It is thus a pure QED object without quarks or gluons and is uncoupled from strong and weak interaction. 

As the electron/positron velocities for bound states are non-relativistic, a fairly precise description of energies and wave functions, sufficient for our cooling model, is provided already by the lowest order expansion of the non-relativistic Dirac equation~\cite{Bohr1947,adkins2018higher}. Here one simply has to enter a reduced mass of $\mu=m_e/2$, where $m_e$ is the mass of the electron, in the corresponding hydrogen expressions. Hence the Bohr-radius increases to the double value of hydrogen: $a_\text{Ps}=2a_0\approx0.11\ \mathrm{nm}$ and the Bohr energy levels read:
\begin{equation}
    E_n=-6.8\ \mathrm{eV}\dfrac{1}{n^2}=\dfrac{E_n^{(H)}}{2},
\end{equation}
where $E_n^{(H)}$ is the energy of the $n$-th energy level of the hydrogen atom. One also gets a much larger de Broglie wavelength of $\lambda_\text{DB} \approx 0.36\ \mu\mathrm{m} $ at $v=1\ \mathrm{km}/\mathrm{s}$ velocity.

This paper investigates laser deceleration and cooling of a hot Ps ensemble in its most simple form of Doppler cooling via velocity dependent radiation pressure. We use combinations of counter-propagating monochromatic Gaussian laser beams and restrict ourselves mostly to proof of principle type fundamental prospects assuming the availability of suitable laser sources. Nevertheless we have to go beyond the oversimplified two-level approximation and work with a more realistic but still truncated sets of energy levels to represent the Ps atom. As we are dealing with triplet ortho-Positronium, Zeeman substructure is present and leads to the appearance of polarization gradient forces~\cite{zimmer2021positronium}.   

\begin{figure}
    \centering
    \includegraphics[width=0.5\linewidth]{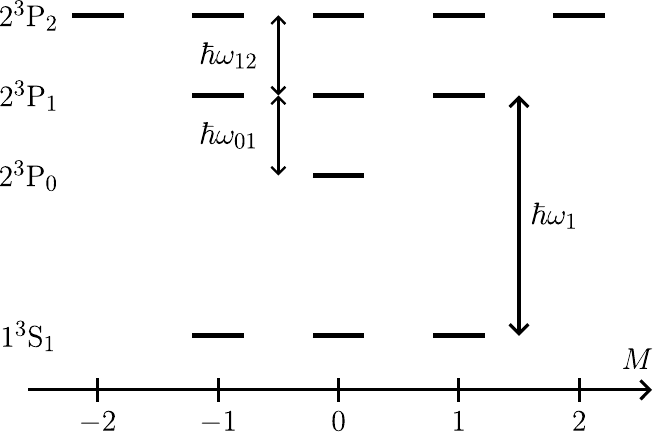}
    \hspace{2cm}
    \includegraphics[width=0.23\linewidth]{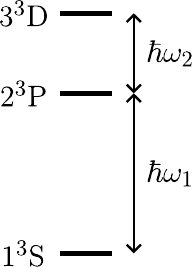}
    \caption{Quantum states of the Zeeman manifold of the triplet Ps $1^3\mathrm{S}$--$2^3\mathrm{P}$ transition used in the 1D simulations (\textit{left}) and the reduced effective 3-level scheme for the 2D simulations of bichromatic cooling (\textit{right})}
    \label{fig:ps-states}
\end{figure}

In our first model targeting the best possible 1D slowing and cooling we include the full Zeeman manifold on the triplet $1^3\mathrm{S}$--$2^3\mathrm{P}$ transition including spin, orbital angular momentum ($\ell$), and projection of their sum ($J$) along the quantization axis ($M$) as depicted in the left side of Figure~\ref{fig:ps-states}. In writing the different states are denoted as $n^3\ell_J$. The upper index $3$ shows that the electron-positron pair is in the triplet spin-state corresponding to ortho-positronium (o-Ps). We only consider this type of Ps as its lifetime ($\approx139\ \mathrm{ns}$) is much longer than the singlet state's one ($\approx0.12\ \mathrm{ns}$). In the left side of Figure~\ref{fig:ps-states} we also indicated the energy difference of some eigenstates. Those with same $n$, $\ell$, and $J$ quantum numbers are considered degenerate.

Applying strong, circularly polarized lasers on this transition will mostly optically polarize the atoms to an effective 2-level scheme on the maximal $m$-manifold. we will make use of this optical pumping in a second model, where we will enlarge our effective model to include an additional transversely applied laser on the $2^3\mathrm{P}$--$3^3\mathrm{D}$ transition. In its simplest form this needs to consider an effective three-level ladder system (see right side of Figure~\ref{fig:ps-states}). Our motivation is two-fold here: populating the $3^3\mathrm{D}$ state with its much smaller annihilation rate can be expected to enhance the Ps lifetime and also lead to improved transverse cooling and confinement.

\subsection{Semi-classical simulation of cooling}

In order to simulate the laser induced light forces and and their application to laser cooling of Ps we have to keep track of its internal quantum state and its external dynamical properties, position and momentum. The total system time evolution is then well described by the Lindblad master equation
\begin{equation}
    \dfrac{\mathrm{d}\rho}{\mathrm{d}t}=-\dfrac{i}{\hbar}\left[H, \rho\right]+\mathcal{L}\rho,
    \label{eq:master}
\end{equation}
where $H$ is the Hamiltonian and $\mathcal{L}\rho$ is the sum of the Liouville damping terms. The latter describes quantum jumps on $N$ possible transitions and reads:
\begin{equation}
    \mathcal{L}\rho=\sum\limits_{k=1}^N\gamma_k\left(2\int N(\vb*{u})J_ke^{-i\vb*{u}\vb*{\hat x}}\rho e^{i\vb*{u}\vb*{\hat x}}J_k^\dagger\, \mathrm{d}\vb*{u}-\left\{J_k^\dagger J_k, \rho\right\}\right).
    \label{eq:liouville}
\end{equation}
Here $J_k$ is the operator of a jump with rate $\gamma_k$. $\vb*{\hat x}$ is the atom's position operator, and $\{\cdot, \cdot\}$ is the anti-commutator of two operators. In this study all quantum jumps represent spontaneous emission of a photon with wave number vector $\vb*{u}$ and a directional distribution of $N(\vb*{u})$.

For this master equation~\eqref{eq:master} no general analytic solution is available and we have to resort to numerical methods. To limit complexity a so called semi-classical description of the atomic motion was adopted in this study. In this well proven semi-classical limit, valid for not too low kinetic temperatures, particle motion is approximated by classical average values of momentum and position, which are well given by their quantum expectation values and their time evolution is governed by classical-type Hamilton-equations of the following form
\begin{equation}
    \label{eq:motion}
    \dfrac{\mathrm{d}\vb*{r}}{\mathrm{d}t}=\dfrac{\vb*{p}}{m}, \quad \dfrac{\mathrm{d}\vb*{p}}{\mathrm{d}t}=\left\langle\vb*{\hat F}\right\rangle_\rho,
\end{equation}
where we introduced the force operator $\vb*{\hat F}=-\vb*{\nabla}H$. Its expectation value for atomic state $\rho$ is denoted by $\left\langle\vb*{\hat F}\right\rangle_\rho$. The atom's mass is $m$.

With these equations we can discard the spatial parts of the Liouville terms~\eqref{eq:liouville}. From now on it will be
\begin{equation}
    \mathcal{L}\rho=\sum\limits_{k=1}^N\gamma_k\left(2J_k\rho J_k^\dagger-\left\{J_k^\dagger J_k, \rho\right\}\right).
\end{equation}
For the complete set of differential equations governing the time evolution of the Ps atoms' motion we add the following equation to~\eqref{eq:motion}.
\begin{equation}
    \begin{aligned}
        \dfrac{\mathrm{d}\rho}{\mathrm{d}t}&=-\dfrac{i}{\hbar}\left[H, \rho\right]+\sum\limits_{k=1}^N\gamma_k\left(2J_k\rho J_k^\dagger-\left\{J_k^\dagger J_k, \rho\right\}\right)
    \end{aligned}
    \label{eq:eq-sys}
\end{equation}

Note that individual random recoil from spontaneous photon emission events is explicitly visible here in the equation for the averages. As recoil is a crucial part of laser cooling counteracting friction forces, we will have to include momentum diffusion explicitly in our individual trajectory simulations to obtain a consistent final temperature. To solve this system of equations~\eqref{eq:motion} and~\eqref{eq:eq-sys} numerically we use the open-source framework \texttt{QuantumOptics.jl}~\cite{kramer2018quantumoptics} in Julia for convenience. It allows to run such semi-classical simulations using stochastic trajectories as they appear in the Monte Carlo wave function simulation (MCWFS) method~\cite{Dum1992, Mlmer1993}. As mentioned above, within the stochastic trajectory method we can include atomic recoil from spontaneous photon emissions during  a quantum jump as an instantaneous change of the momentum by one recoil momentum $\bar k$ pointing in a random direction.

Fortunately for a simple two-level atomic transition in a free plane wave, the average semi-classical force can be analytically calculated to give 
\begin{equation}
F(\Delta, v, \Omega)
= \hbar k\, R_{\rm sc}(\Delta - k v, \Omega)
= \hbar k\, \frac{\Gamma\, \Omega^2}{\Gamma^2 + 2\Omega^2 + 4(\Delta - k v)^2}.
\end{equation} 
or in equivalent form: $
F(\Delta, v)
= \hbar k\, \frac{\Gamma s_0}{2}/({1 + s_0 + \left[{2(\Delta - k v)}\right]^2/ {\Gamma}})$ when rewritten. Using the on-resonance saturation parameter $s_0 = {2\Omega^2}/{\Gamma^2}$. For a strong laser the maximum reduces to $F_\text{max} = \hbar k\, \frac{\Gamma}{2}$ which for the Ps $1^3\mathrm{S}$--$2^3\mathrm{P}$ give a maximum acceleration of $a_\text{max} \approx 10^{12}\ \mathrm{m}/\mathrm{s}^2$ allowing in principle to stop a $50\ \mathrm{km}/\mathrm{s}$ atom in $30\ \mathrm{ns}$. Our goal is to study how close we can get to this limit based on the full Zeeman level scheme in a 3D configuration. Note that this is much shorter than the ground-state annihilation time of $140\ \mathrm{ns}$ so that the majority of the atoms should survive the procedure.   

\subsection{Simulating annihilation}

The major complication and practical challenge in Ps cooling is the atom's finite lifetime against radiative annihilation. Due to angular momentum conservation arguments, the triplet Ortho-positronium ground-state lives much longer ($142\ \mathrm{ns}$) than singlet Ps, as the simplest decay needs already three photons. Hence we concentrate on o-Ps here and only include the $1^3\mathrm{S}$ ground state annihilation as the higher lying states' decay several orders of magnitude slower. Practical simulation of this phenomena was done by including an additional quantum state to those visualised in Figure~\ref{fig:ps-states} - the annihilated state $\ket{a}$. Annihilation of the ground state(s) was treated as quantum jump(s) with a rate of $\gamma_a$. In case of such event the corresponding simulation trajectory was stopped. When an atom is annihilated further keeping track of its motion is irrelevant and thus annihilated atoms were excluded from statistical averages and variations of position and momenta.

\subsection{Radiation pressure cooling including the full Zeeman manifold}

\begin{figure}
    \centering
    \includegraphics[width=\linewidth]{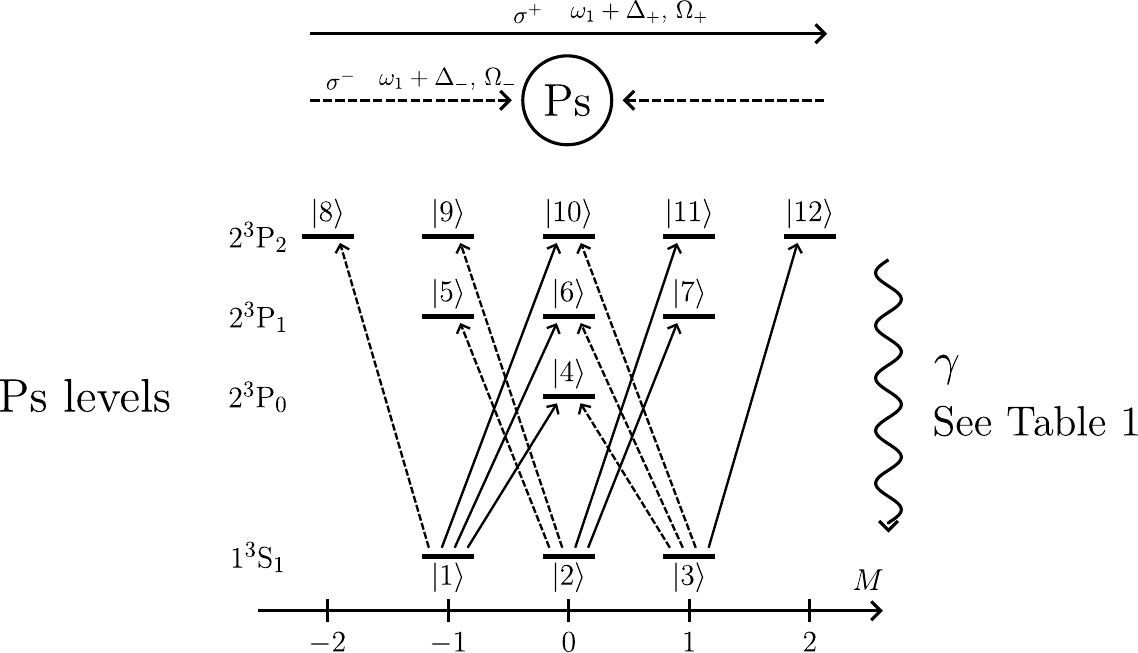}
    \caption{Level couplings between quantum states of the full Zeeman manifold induced by two laser beams along the atomic quantization axis. We assume a strong far red detuned plane running wave laser beam opposing the Ps beam  and a second standing wave field closer to resonance. The first has $\sigma^+$ polarisation, detuning $\Delta_+$ from the transition frequency $\omega_1$ with Rabi frequency $\Omega_+$, while the second is $\sigma^-$ polarized at detuning $\Delta_-$ and amplitude $\Omega_-$.}
    \label{fig:2-level-1d}
\end{figure}

In our first model approach we consider the full Zeeman manifold of quantum states depicted in the left side of Figure~\ref{fig:ps-states} and one-dimensional motion along the cooling laser beams. The setup geometry can be seen schematically in Figure~\ref{fig:2-level-1d} . To reach maximal deceleration and trapping very strong laser fields far beyond saturation power are required, so that all hyperfine upper states are simultaneously coupled. We assume a strong slowing laser beam of $\sigma^+$ polarisation, which excites transitions where the $M$ quantum number increases by $1$ as energy increases. The frequency ($\omega_+$) is far red detuned from the bare atomic transition frequency $\omega_1$ with detuning 
$ \Delta_+=\omega_+-\omega_1 $ and the Rabi frequency generated by the laser electric field is given by
\begin{equation}
    \label{eq:d0}
    \Omega_+=\dfrac{d_0E_+}{\hbar}, \mbox{with} \; d_0=\dfrac{256\sqrt{2}}{243}e\cdot a_0.
\end{equation}
Her $e$ is the elementary charge, and $E_+$ is the amplitude of the oscillating electric field.
In order to achieve better cooling and trapping for low velocities a second beam with polarisation $\sigma^-$ is added, which generates a standing plane wave electric field, coupling transitions $\Delta M = -1$. This beam is characterised by its detuning $\Delta_-$ and Rabi frequency $\Omega_-$.

In Figure~\ref{fig:2-level-1d} all allowed transitions of quantum states are marked with arrows. For simplicity we will use numbers for the different quantum states as shown in the figure.

The Hamiltonian of this system then reads:
\begin{equation}
    \begin{aligned}
        \dfrac{H_1}{\hbar}=&-\Delta_+\ket{7}\bra{7}-\left(\Delta_+-\omega_{12}\right)\left(\ket{11}\bra{11}+\ket{12}\bra{12}\right)-\\
        &-\Delta_-\ket{5}\bra{5}-\left(\Delta_--\omega_{12}\right)\left(\ket{9}\bra{9}+\ket{8}\bra{8}\right)+\\
        &+\dfrac{\Omega_+}{2}\left[e^{-ik_+x}\left(\dfrac{1}{2}\ket{2}\bra{7}+\dfrac{1}{2}\ket{2}\bra{11}+\dfrac{1}{\sqrt{2}}\ket{3}\bra{12}+\right.\right.\\
        &\qquad\left.\left.+e^{i\Delta_+t}\left(\dfrac{1}{2}\ket{1}\bra{6}+\dfrac{e^{-i\omega_{12}t}}{\sqrt{12}}\ket{1}\bra{10}+\dfrac{e^{i\omega_{01}t}}{\sqrt{6}}\ket{1}\bra{4}\right)\right)+\mathrm{H.c.}\right]+\\
        &+\dfrac{\Omega_-}{2}\cos(k_-x)\left[\dfrac{1}{2}\ket{2}\bra{5}+\dfrac{1}{2}\ket{2}\bra{9}+\dfrac{1}{\sqrt{2}}\ket{1}\bra{8}+\right.\\
        &\qquad\left.+e^{i\Delta_-t}\left(\dfrac{1}{2}\ket{3}\bra{6}+\dfrac{e^{-i\omega_{12}t}}{\sqrt{12}}\ket{3}\bra{10}+\dfrac{e^{i\omega_{01}t}}{\sqrt{6}}\ket{3}\bra{4}\right)+\mathrm{H.c.}\right],
    \end{aligned}
\end{equation}
where $k_\pm$ is the wave number of the $\sigma^\pm$ polarised field. For each transition a constant scaling factor was introduced as correction, because transition dipole moments are not equal to $d_0$ defined in~\eqref{eq:d0}, so the corresponding Rabi frequencies differ from $\Omega_\pm$.

For the expectation value of the force we then get:
\begin{equation}
    \begin{aligned}
        \dfrac{\left\langle\hat F_1\right\rangle_\rho}{\hbar}=&-k_+\Omega_+\mathrm{Im}\left[e^{-ik_+x}\left(\dfrac{1}{2}\left\langle\ket{2}\bra{7}\right\rangle_\rho+\dfrac{1}{2}\left\langle\ket{2}\bra{11}\right\rangle_\rho+\dfrac{1}{\sqrt{2}}\left\langle\ket{3}\bra{12}\right\rangle_\rho+\right.\right.\\
        &\qquad\left.\left.+e^{i\Delta_+t}\left(\dfrac{1}{2}\left\langle\ket{1}\bra{6}\right\rangle_\rho+\dfrac{e^{-i\omega_{12}t}}{\sqrt{12}}\left\langle\ket{1}\bra{10}\right\rangle_\rho+\dfrac{e^{i\omega_{01}}}{\sqrt{6}}\left\langle\ket{1}\bra{4}\right\rangle_\rho\right)\right)\right]+\\
        &+k_-\Omega_-\sin(k_-x)\mathrm{Re}\left[\dfrac{1}{2}\left\langle\ket{2}\bra{5}\right\rangle_\rho+\dfrac{1}{2}\left\langle\ket{2}\bra{9}\right\rangle_\rho+\dfrac{1}{\sqrt{2}}\left\langle\ket{1}\bra{8}\right\rangle_\rho+\right.\\
        &\qquad\left.+e^{i\Delta_-t}\left(\dfrac{1}{2}\left\langle\ket{3}\bra{6}\right\rangle_\rho+\dfrac{e^{-i\omega_{12}t}}{\sqrt{12}}\left\langle\ket{3}\bra{10}\right\rangle_\rho+\dfrac{e^{i\omega_{01}}}{\sqrt{6}}\left\langle\ket{3}\bra{4}\right\rangle_\rho\right)\right].
    \end{aligned}
\end{equation}
 Here $\mathrm{Re}[z]$ and $\mathrm{Im}[z]$ denote the real-, and imaginary parts of $z\in\mathbb{C}$ respectively.

To take spontaneous emission events into account we introduce the $J_k$ jump operators as regular transition operators $\ket{i}\bra{j}$. The corresponding rates are summarised in Table~\ref{tab:rates} in units of $\gamma_0$ i.e., the $\ket{1}\bra{8}$ jump operator's corresponding rate, which is $\gamma_0\approx315\ \mathrm{MHz}$.

To take the recoil caused by the emitted photon into consideration the atoms' momentum was changed instantaneously by the emitted photon's momentum in a random direction every time such event occurred. The absolute value of this recoil is determined as
\begin{equation}
    p_\text{recoil}=\omega_\text{photon}\dfrac{\hbar}{c} = \hbar k_\text{photon},
\end{equation}
where $\omega_\text{photon}\in\{\omega_1-\omega_{01}, \omega_1, \omega_1+\omega_{12}\}$ is the emitted photon's frequency and $k_\text{photon}$ is the corresponding wave number.

\begin{table}
    \centering
    \begin{tabular}{|c|c|c|c|c|c|c|c|c|c|}
    \hline
    $\ket{i}\bra{j}$ & $\bra{4}$ & $\bra{5}$ & $\bra{6}$ & $\bra{7}$ & $\bra{8}$ & $\bra{9}$ & $\bra{10}$ & $\bra{11}$ & $\bra{12}$ \\ \hline
    $\ket{1}$        & $1/3$     & $0$       & $1/2$     & $1/2$     & $1$       & $1/2$     & $1/6$      & $0$        & $0$        \\
    $\ket{2}$        & $1/3$     & $1/2$     & $0$       & $1/2$     & $0$       & $1/2$     & $2/3$      & $1/2$      & $0$        \\
    $\ket{3}$        & $1/3$     & $1/2$     & $1/2$     & $0$       & $0$       & $0$       & $1/6$      & $1/2$      & $1$        \\ \hline
    \end{tabular}
    \caption{Spontaneous emission rates corresponding to the jump operators $\ket{i}\bra{j}$ in units of $\gamma_0$.}
    \label{tab:rates}
\end{table}

As discussed above annihilation of the ground states is also implemented via quantum jumps to an auxiliary level with jump operators
\begin{equation}
    \ket{a}\bra{1}, \qquad \ket{a}\bra{2}, \qquad \ket{a}\bra{3}
\end{equation}
and rate
\begin{equation}
    \gamma_a\approx7.04\ \mathrm{MHz}.
\end{equation}
When such an event occurs the further simulation of this trajectory is stopped. As discussed in the results section below the central goal is to find optimal parameters yielding high phase space density ensemble with as many atoms as possible at low temperature. 

\subsection{Bichromatic cooling using an effective 3-level ladder model}

\begin{figure} 
    \centering
    \includegraphics[width=0.7\textwidth]{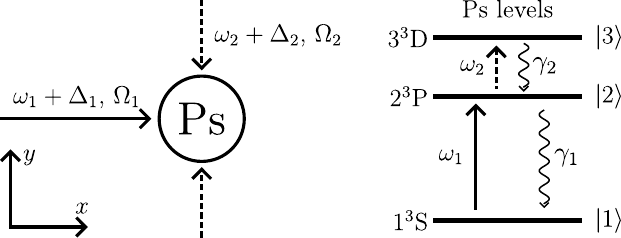}
    \caption{Geometry of the effective bichromatic three-level 2D cooling scheme with a Gaussian slowing beam of width $w_1$ pointing against the Ps source at detuning $\Delta_1$ and with Rabi frequency $\Omega_1$ on the lower transition and a transverse plane standing wave cooling laser at detuning $\Delta_2$ from frequency from the upper transition with amplitude $\Omega_2$.}
    \label{fig:3-level-2d}
\end{figure}

In our second model approach we target possible improvements of Ps cooling by involving higher Ps states with longer annihilation lifetimes. As a generic and not too complex example we consider a reduced three-level model as depicted on the right side of Figure~\ref{fig:ps-states} and also include motion in longitudinal and transverse direction. The cooling setup can be seen in Figure~\ref{fig:3-level-2d}. Here we examined the application of a running wave slowing laser exciting the $1^3\mathrm{S}$--$2^3\mathrm{P}$ transition and a standing wave transverse cooling laser on the $2^3\mathrm{P}$--$3^3\mathrm{D}$ one. The former has detuning from transition frequency $\omega_1$ of $\Delta_1$ and strength $\Omega_1$. Laser two is detuned from $\omega_2$ by $\Delta_2$ and has a strength of $\Omega_2$. In the two-dimensional case we assume laser beams with a Gaussian profile in the direction perpendicular to their wave vector. In this scheme the running wave's profile had a width of $w_1=1000\lambda_1=1000\cdot2\pi c/\omega_1$, while the transverse beam was considered very large: $w_2\to\infty$. \\
\vspace{1cm}

The corresponding Hamiltonian in a rotating frame reads:
\begin{equation}
    \dfrac{H_2}{\hbar}=\Delta_1\ket{1}\bra{1}-\Delta_2\ket{3}\bra{3}+\left[\dfrac{\Omega_1}{2}e^{-\frac{y^2}{w_1^2}-ik_1x}\ket{1}\bra{2}+\dfrac{\Omega_2}{2}\cos\left(k_2y\right)\ket{2}\bra{3}+\mathrm{H.c.}\right],
\end{equation}
where $k_1$ and $k_2$ are wave numbers of the running and standing waves respectively. Expectation values of the force operator in the $x$-, and $y$ directions are
\begin{equation}
    \begin{aligned}
        \dfrac{\left\langle\hat F_{2, x}\right\rangle_\rho}{\hbar}=&-\Omega_1k_1e^{-\frac{y^2}{w_1^2}}\mathrm{Im}\left[e^{-ik_1x}\left\langle\ket{1}\bra{2}\right\rangle\right],\\
        \dfrac{\left\langle\hat F_{2, y}\right\rangle_\rho}{\hbar}=&\dfrac{2\Omega_1}{w_1^2}ye^{-\frac{y^2}{w_1^2}}\mathrm{Re}\left[e^{-ik_1x}\left\langle\ket{1}\bra{2}\right\rangle\right]+\Omega_2k_2\sin\left(k_2y\right)\mathrm{Re}\left[\left\langle\ket{2}\bra{3}\right\rangle\right]
    \end{aligned}
\end{equation}
with jump operators given by
\begin{equation}
    \hat{J}_{12} =\ket{1}\bra{2} \quad \text{and } \hat{J}_{23} = \ket{2}\bra{3},
\end{equation}
with spontaneous emission rates $\gamma_1$ and $\gamma_2$. At each jump event an instantaneous momentum kick is added to the atom's momentum in a random direction with absolute value $ p_\text{recoil}=\omega_\text{photon}\dfrac{\hbar}{c} $. Here $\omega_\text{photon}\in\{\omega_1, \omega_2\}$ is the emitted photon's frequency. Annihilation of the Ps atoms is modelled analogously with a jump operator $\ket{a}\bra{1}$ and rate $\gamma_a$.

\subsection{Two step slowing and trapping using an additional standing wave}
\label{running-standing}

In order to further optimize the cooling process, we analysed a modified version of this two-dimensional scheme. After slowing down the atoms in the $x$-direction for some time $t_r$ with a running wave laser, illumination is changed to a standing wave to induce better cooling and create an optical trapping potential. Experimentally this amounts to splitting the power of the laser into two counter-propagating running waves of equal intensity and frequency, which changes the Hamiltonian to:
\begin{equation}
    \dfrac{H_2'}{\hbar}=\Delta_1\ket{1}\bra{1}-\Delta_2\ket{3}\bra{3}+\left[\dfrac{\sqrt{2}}{2}\Omega_1e^{-\frac{y^2}{w_1^2}}\cos\left(k_1x\right)\ket{1}\bra{2}+\dfrac{\Omega_2}{2}\cos\left(k_2y\right)\ket{2}\bra{3}+\mathrm{H.c.}\right],
\end{equation}
and the force operator's expectation values are
\begin{equation}
    \begin{aligned}
        \dfrac{\left\langle\hat F_{2, x}'\right\rangle_\rho}{\hbar}=&\sqrt{2}\Omega_1k_1e^{-\frac{y^2}{w_1^2}}\sin\left(k_1x\right)\mathrm{Re}\left[\left\langle\ket{1}\bra{2}\right\rangle\right],\\
        \dfrac{\left\langle\hat F_{2, y}'\right\rangle_\rho}{\hbar}=&\dfrac{2\sqrt{2}\Omega_1}{w_1^2}ye^{-\frac{y^2}{w_1^2}}\cos\left(k_1x\right)\mathrm{Re}\left[\left\langle\ket{1}\bra{2}\right\rangle\right]+\Omega_2k_2\sin\left(k_2y\right)\mathrm{Re}\left[\left\langle\ket{2}\bra{3}\right\rangle\right].
    \end{aligned}
\end{equation}
The central goal of this simulation is to highlight the possible benefits of bichromatic cooling in a ladder configuration involving higher excited Ps states.  

\section{Simulation Results and Discussion}
\label{results}

\textbf{Initial conditions:} Before presenting the central results of our simulations for the two above geometries, we have to define our initial ensemble. As mentioned before, our primary Ps source are ns-long bunches created from positron pulses send to a silica micro channel trap, which constitutes the current state of the art Ps source~\cite{mariazzi2021high}. Hence we start from a submillimetre localized hot cloud moving away from the target surface towards the incoming slowing beam. For simplicity we assume a 3D Gaussian momentum distribution without backward momenta.
For each individual parameter set we simulated $200$ independent trajectories with momenta picked randomly from the initial momentum distribution. We used a reduced transverse momentum cut-off to select atoms moving away from the surface at angles not too far from $90^\circ$. We start from a cloud with an average momentum of the order of $\bar{p} \approx 10\, p_\text{rec}$ i.e., $v\approx 15\ \mathrm{km}/\mathrm{s}$, which is slower than emitted from current room temperature targets, but could be reached from colder target surfaces.  

The trajectories were the basis of our statistical analysis. In every case the Ps atom's initial position was set to the origin of the used coordinate system. Their initial momentum was chosen randomly according to normal distributions with probability density functions
\begin{equation}
    \begin{aligned}
        \varrho_x(p_x)&=\dfrac{1}{\sqrt{2\pi\sigma_x^2}}\exp\left(-\dfrac{(p_x-p_{x, 0})^2}{2\sigma_x^2}\right),\\
        \varrho_y(p_y)&=\dfrac{1}{\sqrt{2\pi\sigma_y^2}}\exp\left(-\dfrac{(p_y-p_{y, 0})^2}{2\sigma_y^2}\right)
    \end{aligned}
\end{equation}
with mean values and standard deviations of
\begin{equation}
    \begin{aligned}
        p_{x, 0}&=-10\, \omega_1/c, &\sigma_x^2&=4\, \omega_1/c,\\
        p_{y, 0}&=0, &\sigma_y^2&=0.5\, \omega_1/c.
    \end{aligned}
\end{equation}
These values are realistic for Ps generating. For the one-dimensional case only $\varrho_x(p_x)$, for the two-dimensional one both $\varrho_x(p_x)$ and $\varrho_y(p_y)$ were applied.

The atoms' initial quantum state was set to $\ket{2}$ for the 1D case (Figure~\ref{fig:2-level-1d}) and $\ket{1}$ for the 2D one (Figure~\ref{fig:3-level-2d}). These choices are reasonable because after their formation - but before laser beams are turned on - excited atoms quickly decay to the $1^3\mathrm{S}$ ground state.


In the following subsections we present the key simulation results obtained by covering a large range of parameter sets for which we determined the atomic positions, momenta, populations and an effective temperature in relation to their kinetic energy spread. For the concrete examples we selected special parameter combinations which yielded  
\begin{enumerate}

    \item the lowest temperature,
    \item the highest number of non-annihilated atoms, and
    \item the highest number of non-annihilated atoms at the time when temperature reached $1\ \mathrm{K}$.

\end{enumerate}

\subsection{Plane wave 1D radiation pressure deceleration and cooling on full Zeeman manifold}

In order to find good parameters for the one-dimensional cooling setup we examined $625$ different values for the $\{\Delta_+, \Delta_-, \Omega_+, \Omega_-\}$ set of parameters. Each element of this set took five values from the following intervals evenly distributed:
\begin{itemize}

    \item $\Delta_+\in\left[-430\ \mathrm{GHz}; -330\ \mathrm{GHz}\right]$,
    \item $\Delta_-\in\left[-100\ \mathrm{GHz}; 0\ \mathrm{GHz}\right]$,
    \item $\Omega_+\in\left[600\ \mathrm{GHz}; 1000\ \mathrm{GHz}\right]$, and
    \item $\Omega_-\in\left[600\ \mathrm{GHz}; 1000\ \mathrm{GHz}\right]$.
    
\end{itemize}

These values were chosen because of some physical considerations and experimental limitations. The running wave with $\sigma^+$ polarisation is used for slowing down the initially fast atoms. For that the laser frequency observed by the moving atom has to be close to the transition frequency, that is why $\Delta_+$ detuning is large. The other laser beam's detuning ($\Delta_-$) is smaller, because it is used for the trapping of the atoms already slowed down. Rabi frequency of both beams is maximised at $1000\ \mathrm{GHz}$. This is because the corresponding field strength requires laser intensity of around $10^7\ \mathrm{W}/\mathrm{cm}^2$, which is the maximum for ordinary lasers in laboratories.

To determine the temperature of the atomic ensemble we used the variation of the non-annihilated atoms' momentum. Using the one-dimensional Maxwell--Boltzmann distribution we obtain temperature as
\begin{equation}
    \label{eq:1d-temp-def}
    T=\dfrac{\mathrm{Var}(p)}{m\cdot k_B},
\end{equation}
where $\mathrm{Var}(p)=\left\langle p^2\right\rangle-\left\langle p\right\rangle^2$ is the variation of momentum and $k_B$ is the Boltzmann constant. This is a kinetic definition of temperature depending only on the motion of the atoms, which does not take their internal energy and quantum state into account.

\begin{figure}
    \centering
    \includegraphics[width=0.97\linewidth]{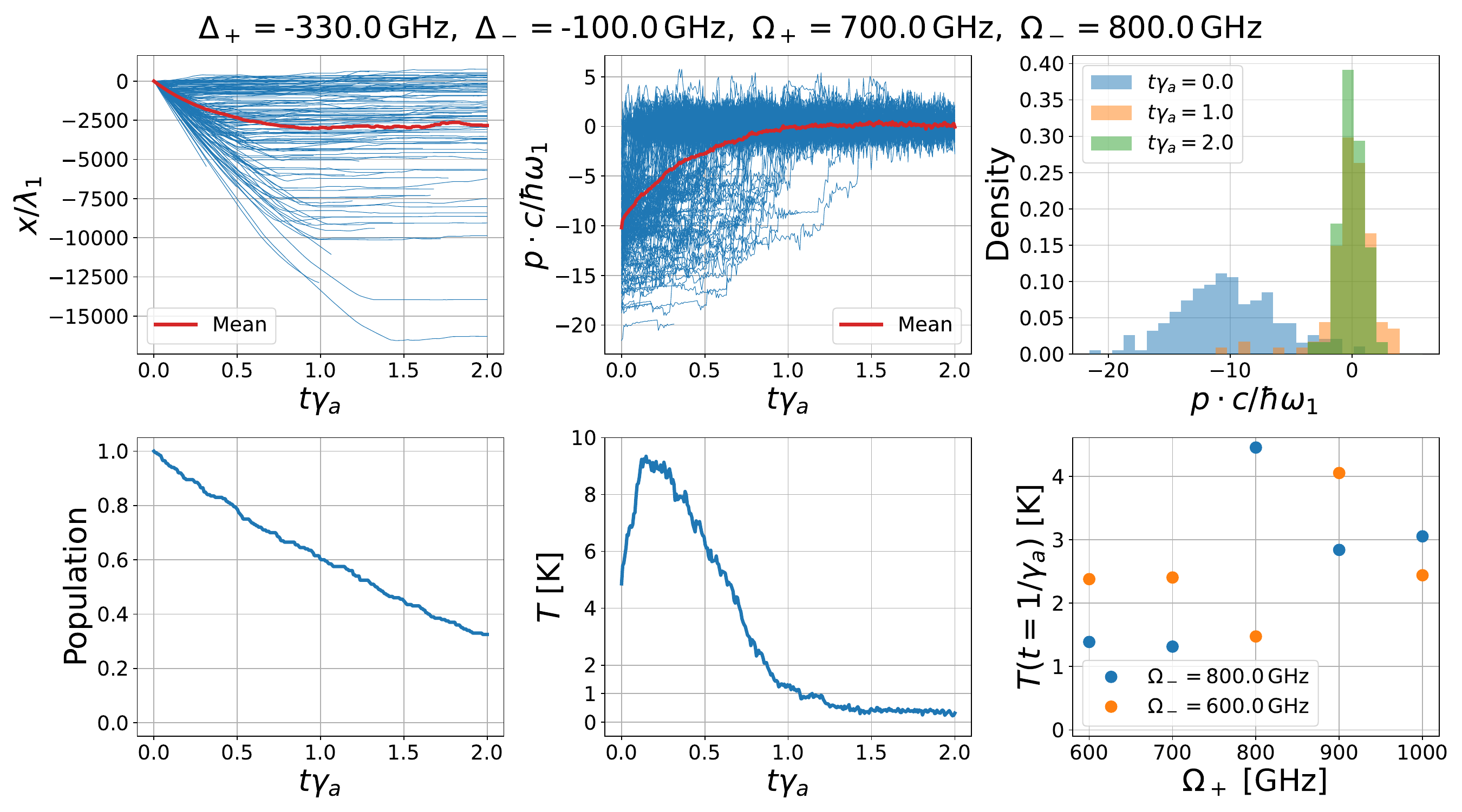}
    \caption{Time evolution of atomic positions (\textit{top left}), momenta (\textit{top centre}), the momentum distribution (\textit{top right}), fraction of non-annihilated atoms (\textit{bottom left}), and temperature as defined in Eq.~\eqref{eq:1d-temp-def} (\textit{bottom centre}) for laser parameters from Eq.~\eqref{eq:1d-temp}. This choice resulted in the lowest achieved temperature ($0.22\ \mathrm{K}$) for the one-dimensional case. We also highlighted the temperature after one lifetime (\textit{bottom right}) for various $\Omega_+$ and $\Omega_-$ values while keeping the detunings the same.}
    \label{fig:1d-temp}
\end{figure}

With this definition of temperature we achieved the lowest value of $T\approx 0.22K$ for parameters:
\begin{equation}
    \label{eq:1d-temp}
    \Delta_+=-330\ \mathrm{GHz}, \quad \Delta_-=-100\ \mathrm{GHz}, \quad \Omega_+=700\ \mathrm{GHz}, \quad \Omega_-=800\ \mathrm{GHz}.
\end{equation}
Time evolution of each trajectory and statistical variables can be seen in Figure~\ref{fig:1d-temp}. Here we plotted each atom's position and momentum as a function of time in units of $\lambda_1=2\pi c/\omega_1\approx1.5\ \mu\mathrm{m}$ and $\hbar\omega_1/c$ respectively. Mean value of these quantities are also shown with red line. The cooling and trapping process is clear from the graphs. Most of the atoms have not moved more than $10^5\lambda_1\approx15\ \mathrm{mm}$. In the top right corner of Figure~\ref{fig:1d-temp} we plotted the distribution of the non-annihilated atoms' momentum at the beginning of the process, after one, and two lifetimes ($1/\gamma_a$). It is easy to see that after only one lifetime most Ps atoms' momentum oscillates around $0$, which is an indicator of trapping and at the end all of the non-annihilated atoms got trapped.

On the bottom centre graph one can see the temperature calculated for the ensemble of the non-annihilated atoms. We can see a fast cooling from $9\ \mathrm{K}$ to $1\ \mathrm{K}$ under one lifetime and it cooled down further to $0.22\ \mathrm{K}$, which was the coldest ensemble found for this process geometry. Note that we get close to the recoil temperature, but still are stuck at least one order of magnitude above the Doppler temperature due to the large recoil heating. 
For these parameters the surviving atom number decreases rather quickly. After one lifetime $40\%$, after two lifetimes more than $60\%$ of the atoms got annihilated. To show the efficiency of cooling we plotted the temperature at time $t=1/\gamma_a$ for different $\Omega_+$ and $\Omega_-$ Rabi frequencies. This illustrates a complex dependence of the slowing and cooling rate on laser powers, which can be optimized according to the desired ensemble properties. 

\begin{figure}
    \centering
    \includegraphics[width=\linewidth]{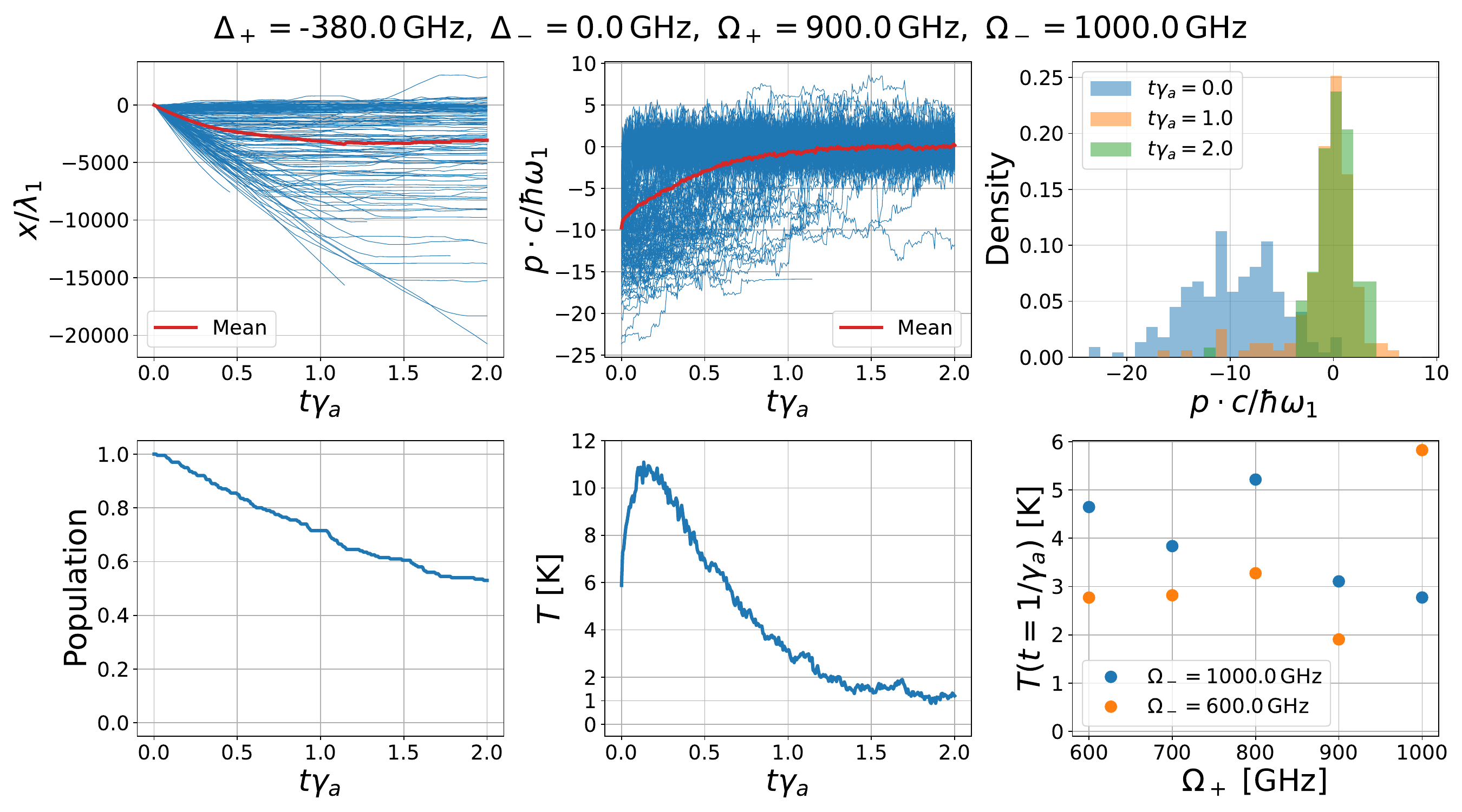}
    \caption{Same as in the above Figure~\ref{fig:1d-temp} with laser parameters~\eqref{eq:1d-pop}. These parameters yield the highest number of non-annihilated cold atoms at the end of the simulation ($53\%$).}
    \label{fig:1d-pop}
\end{figure}

Using alternative parameters
\begin{equation}
    \label{eq:1d-pop}
    \Delta_+=-380\ \mathrm{GHz}, \quad \Delta_-=0\ \mathrm{GHz}, \quad \Omega_+=900\ \mathrm{GHz}, \quad \Omega_-=1000\ \mathrm{GHz}
\end{equation}
we predict the highest number of non-annihilated atoms at the end of the process at somewhat higher temperature. Figure~\ref{fig:1d-pop} shows the cooling procedure similarly to the previous case. Trapping is clear in this one too, but we can see that the atoms get slightly further from their initial position. There are also more atoms that kept a large momentum, but most of them still got slowed down greatly.

It is visible that only half of the atoms got annihilated during the two lifetimes long process. Unfortunately temperature decreased rather slowly. It reached $1\ \mathrm{K}$ only at the very end of the process. Considering that less than half of the atoms got annihilated during this time it is still a great result.

\begin{figure}
    \centering
    \includegraphics[width=\linewidth]{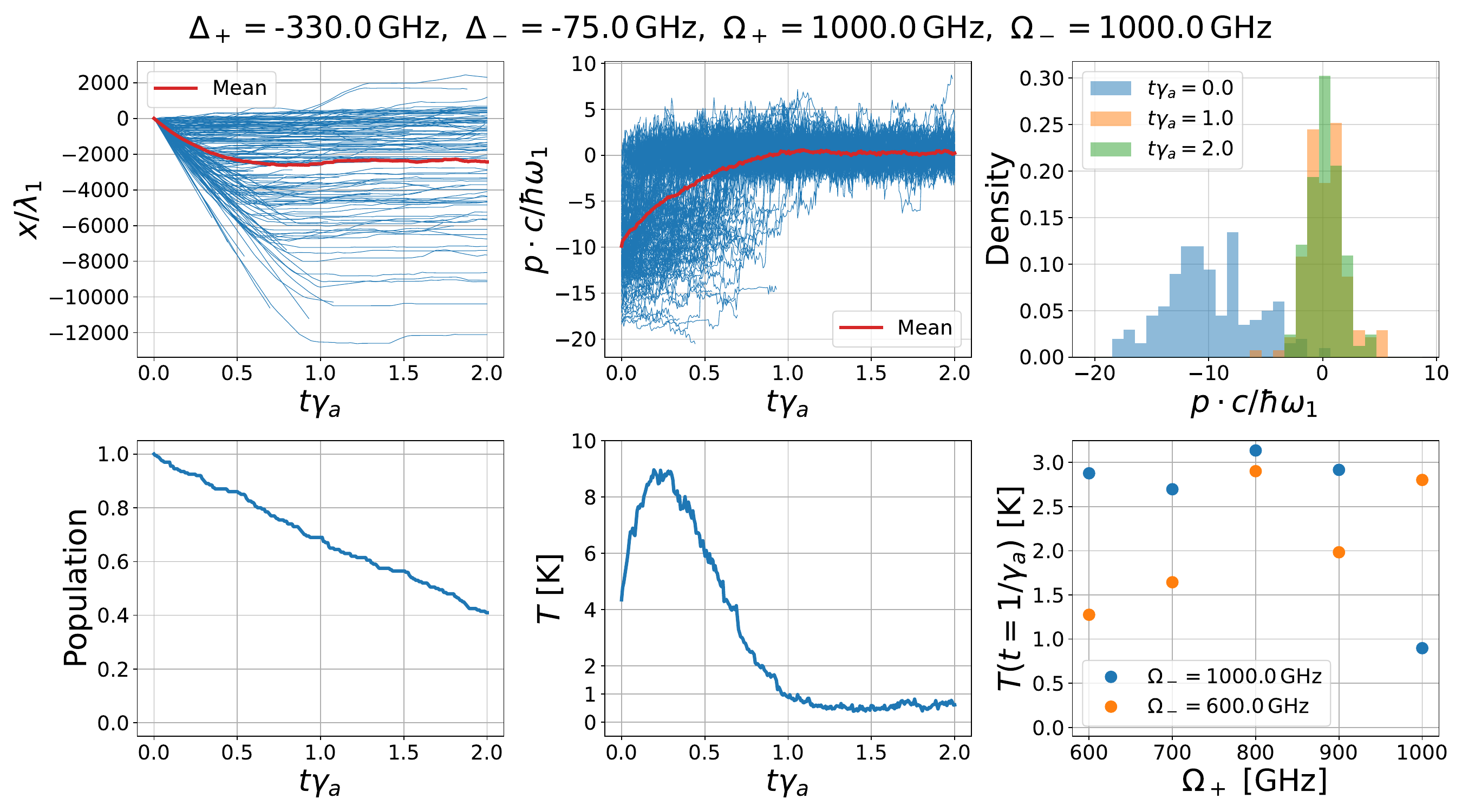}
    \caption{Position and momentum time evolution as in Figure~\ref{fig:1d-temp} during the cooling pulse with laser parameters~\eqref{eq:1d-mix}. These parameters yield the highest number of surviving Ps atoms at a temperature of $T < 1\ \mathrm{K}$ ($69\%$) for the one-dimensional case.}
    \label{fig:1d-mix}
\end{figure}

Moreover, we also looked at another quality measure: the number of non-annihilated atoms at the time of reaching $1\ \mathrm{K}$ temperature. According to this one the parameters
\begin{equation}
    \label{eq:1d-mix}
    \Delta_+=-330\ \mathrm{GHz}, \quad \Delta_-=-75\ \mathrm{GHz}, \quad \Omega_+=1000\ \mathrm{GHz}, \quad \Omega_-=1000\ \mathrm{GHz}
\end{equation}
provided the best results. We summarised them in Figure~\ref{fig:1d-mix}. In this case too most atoms got trapped within $15\ \mathrm{mm}$ from their initial position. Their slowing was also fast, after just one lifetime almost all of them had momentum less than $5\hbar\omega_1/c$ in absolute value.

The share of atoms that did not annihilate was promising as around $40\%$ of the atoms were preserved. One can see a rapid decreasing in the temperature too similar to the first case. The ensemble of non-annihilated atoms reached a temperature of $1\ \mathrm{K}$ under one lifetime and dropped further afterwards to $0.4\ \mathrm{K}$. When it cooled down to $1\ \mathrm{K}$ $69\%$ of all atoms were still in the atomic states.

In summary we confirm earlier predictions of the efficiency of laser cooling of Ps allowing to reach sub-Kelvin temperatures down to about $T_\text{rec}\approx0.15\ \mathrm{K}$ in approximately $200\ \mathrm{ns}$ keeping up to half of the Ps atoms. Significantly faster cooling to about $1\ \mathrm{K}$ is possible with stronger laser amplitudes, corresponding to Rabi frequencies of around $\approx 3000\gamma_0$. While this sounds very high, it is within reach of current technology for focussed $\mathrm{mJ}$-energy sub $100\ \mathrm{ns}$ pulses. 

\subsection{2D cooling on effective 3-level ladder model}
Now let us turn our attention to including a third higher up energy level and change to a two-dimensional model. The parameter scan was conducted similarly to the previous case. The values for the $\{\Delta_1, \Delta_2, \Omega_1, \Omega_2\}$ set of parameters were chosen as five-five evenly distributed points from the intervals
\begin{itemize}

    \item $\Delta_1\in\left[-430\ \mathrm{GHz}; -330\ \mathrm{GHz}\right]$,
    \item $\Delta_2\in\left[-100\ \mathrm{GHz}; 0\ \mathrm{GHz}\right]$,
    \item $\Omega_1\in\left[600\ \mathrm{GHz}; 1000\ \mathrm{GHz}\right]$, and
    \item $\Omega_2\in\left[600\ \mathrm{GHz}; 1000\ \mathrm{GHz}\right]$.
    
\end{itemize}

The running wave laser detuning $\Delta_1$ is chosen in a way that the power broadened line covers the major fraction of the initial velocity distribution allowing it to slow many atoms down in the $x$-direction. The second laser on the upper transition perpendicular to this motion will reduce the fast annihilating ground state population ($\ket{1}$) and additionally adds transverse cooling. we consider atoms which do not have large initial momentum components in the   $y$-direction so that $\Delta_2$ can be chosen much smaller not so large. The corresponding standing wave serves as trapping potential in the $y$-direction too. Rabi frequencies again are maximal at $1000\ \mathrm{GHz}$, so that the required laser intensities do not exceed $10^7\ \mathrm{W}/\mathrm{cm}^2$, which corresponds to $10\ \mathrm{mJ}$ of pulse energy for a duration of $100\ \mathrm{ns}$ when focussed to an area of $1\ \mathrm{mm}^2$.

Here temperature was determined via fitting to a two-dimensional Maxwell-Boltzmann distribution
\begin{equation}
    \label{eq:2d-temp-def}
    \rho_\text{MB}(|\vb*{p}|)=\dfrac{|\vb*{p}|}{mk_BT}\cdot\exp\left(-\dfrac{|\vb*{p}|^2}{2mk_BT}\right).
\end{equation}
to the atomic momenta  with fitting parameter $T$. This is a kinetic energy based temperature definition which does not take the atoms' internal energy and quantum state into account. It is known not be a perfect fit for Doppler cooling but still gives a solid and reliable estimate.

\begin{figure}
    \centering
    \includegraphics[width=0.96\linewidth]{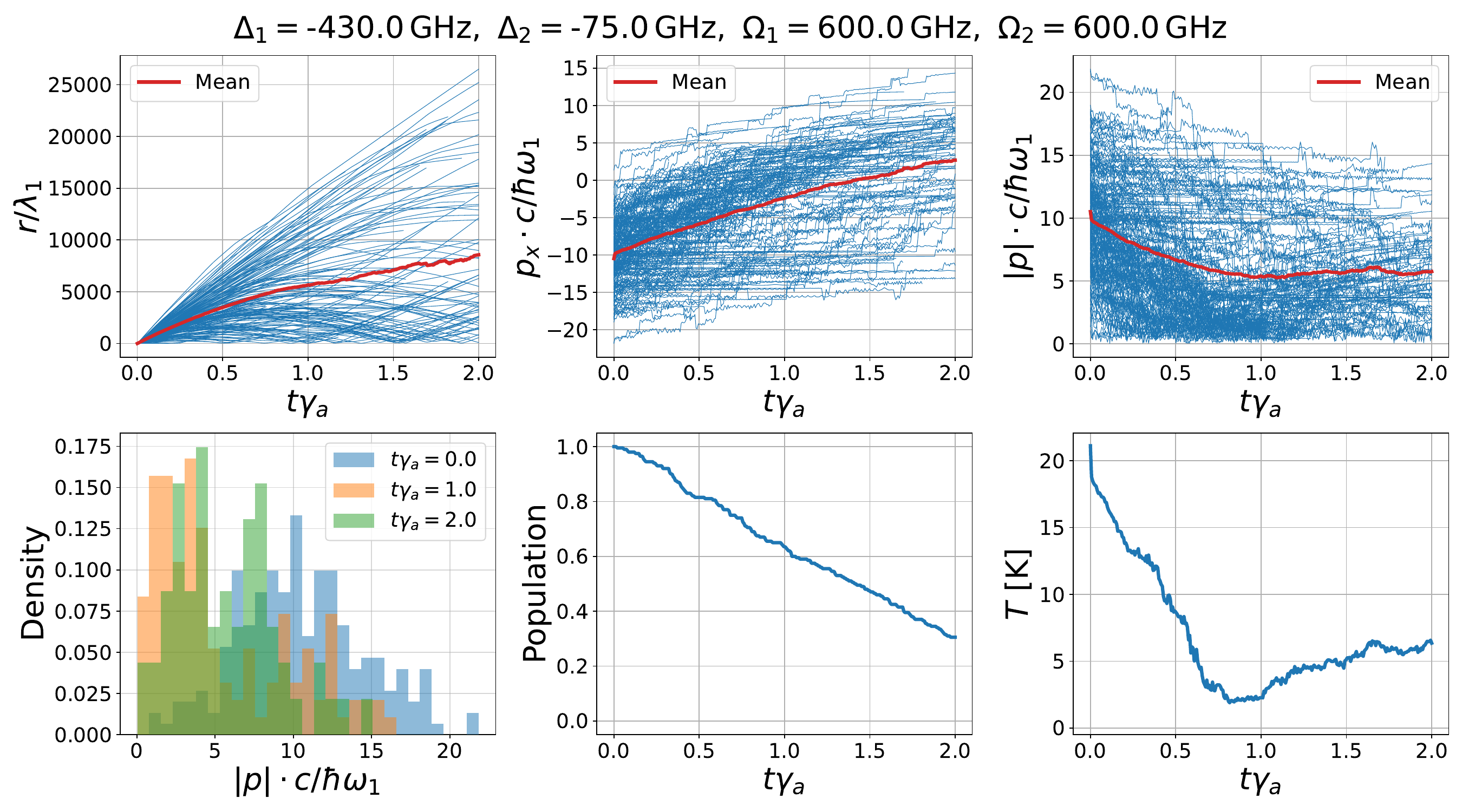}
    \caption{Time evolution of atomic distance from the origin (\textit{top left}), momentum in the $x$-direction (\textit{top centre}), absolute value of momentum (\textit{top right}), its distribution (\textit{bottom left}), population of non-annihilated atoms (\textit{bottom centre}), and fitted temperature estimate of~\eqref{eq:2d-temp-def} (\textit{bottom right}), with laser parameters~\eqref{eq:2d-temp}. The choice resulted in the lowest achieved temperature ($1.88\ \mathrm{K}$) for the two-dimensional case.}
    \label{fig:2d-temp}
\end{figure}

The lowest temperature was achieved with the parameter set
\begin{equation}
    \label{eq:2d-temp}
    \Delta_1=-430\ \mathrm{GHz}, \quad \Delta_2=-75\ \mathrm{GHz}, \quad \Omega_1=600\ \mathrm{GHz}, \quad \Omega_2=600\ \mathrm{GHz}.
\end{equation}
In this case $1.88\ \mathrm{K}$ was reached in under one lifetime as it can be seen in Figure~\ref{fig:2d-temp}. Here we also plotted the distance of the atoms from their initial position (top left) and their momentum in the $x$-direction as a function of time. We can see that some of them got as far as $25,000\lambda_1\approx37.5\ \mathrm{mm}$ from the origin. It is also notable that in the $x$-direction momentum decreases linearly, but does not stop at $0$. This means that after a certain time atoms start re-accelerating in the laser beam's direction and thus increasing the ensemble's temperature. This phenomena is visible on the other plots too showing the time evolution of the atoms' momentum and its distribution. During this process a large amount of atoms got annihilated quite quickly. Only $30\%$ of them remained in atomic form after two lifetimes.

\begin{figure}
    \centering
    \includegraphics[width=0.96\linewidth]{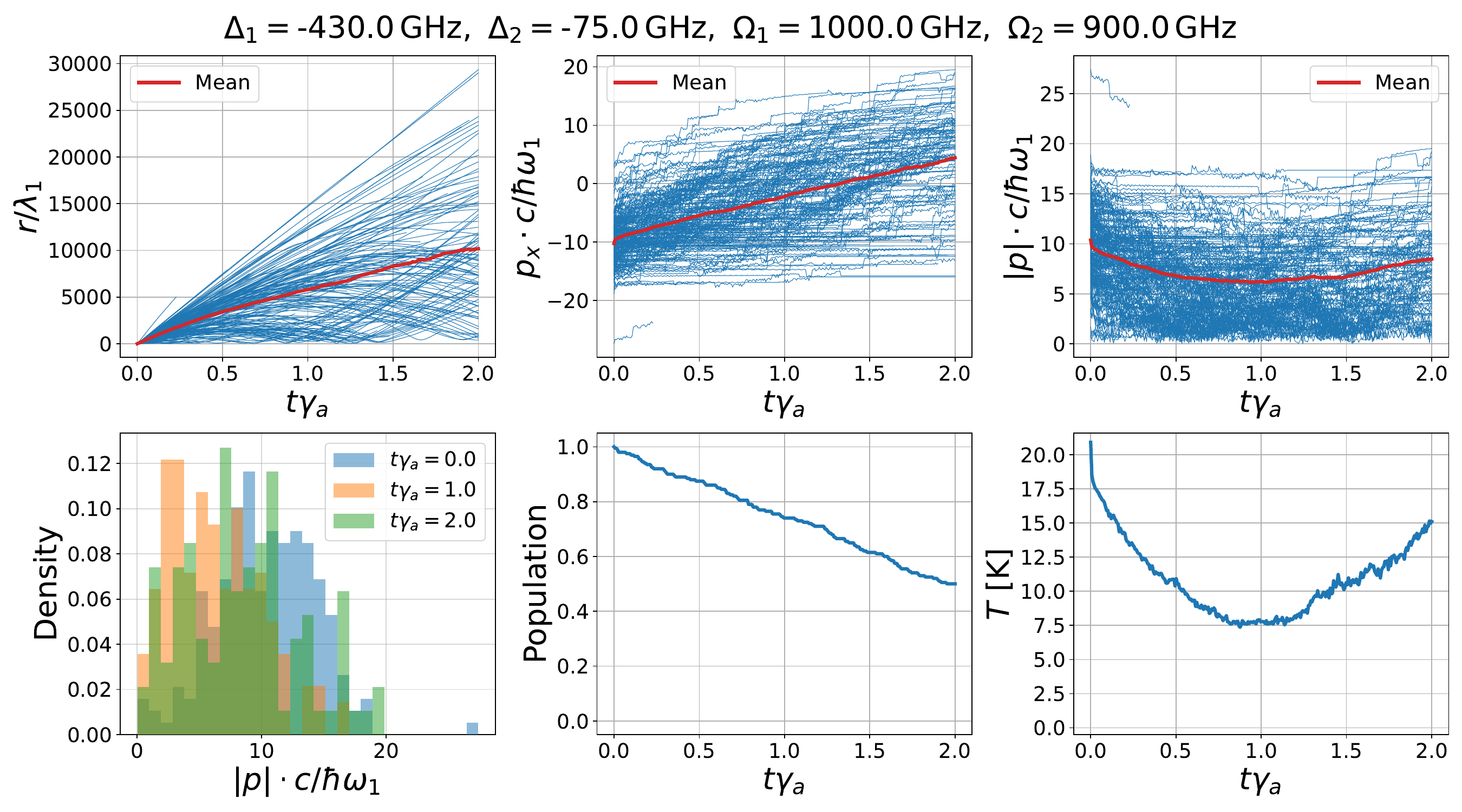}
    \caption{Same as in Figure~\ref{fig:2d-temp} with laser parameters~\eqref{eq:2d-pop} yielding the highest number of non-annihilated atoms at the end of the simulation ($50\%$) for the two-dimensional case.}
    \label{fig:2d-pop}
\end{figure}

The most Ps atoms were preserved for applying
\begin{equation}
    \label{eq:2d-pop}
    \Delta_1=-430\ \mathrm{GHz}, \quad \Delta_2=-75\ \mathrm{GHz}, \quad \Omega_1=1000\ \mathrm{GHz}, \quad \Omega_2=900\ \mathrm{GHz}.
\end{equation}
In this case half of the atoms got annihilated after two lifetimes, but as it is visible in Figure~\ref{fig:2d-pop} no effective cooling was achieved. The biggest problem we have to tackle is the running wave's re-accelerating property.

To come over this issue one can switch from running wave electric field to a standing one after a given time $t_r$ as it is described in subsection~\ref{running-standing}. This way we can harness the slowing effect of the running wave and the trapping of the standing one when the atoms are moving slower.

To run simulations for this case we chose to use the same laser parameters as described in the beginning of this subsection. The time of switching the running wave electric field to a standing one was set to $t_r=0.75/\gamma_a$. We chose this value because in the best case (Figure~\ref{fig:2d-temp}) this was the time needed to reach the lowest temperature with a constant running wave in the $x$-direction.

\begin{figure}
    \centering
    \includegraphics[width=\linewidth]{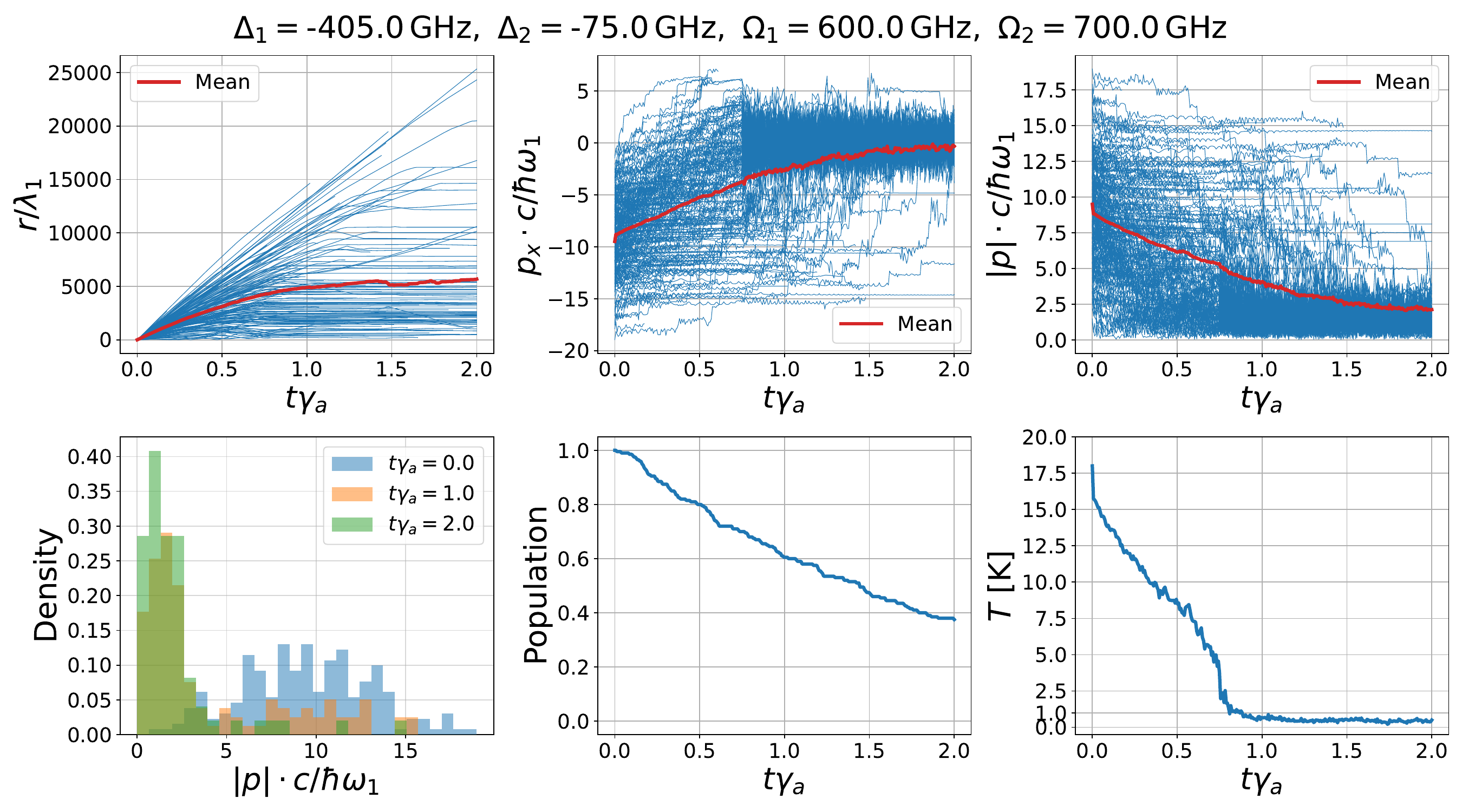}
    \caption{Same as in Figure~\ref{fig:2d-temp}, with laser parameters~\eqref{eq:2d-temp-change}. This resulted in the lowest achieved temperature ($0.22\ \mathrm{K}$) for the two-dimensional case, when the running wave electric field was changed to a standing one after time $t_r=0.75/\gamma_a$.}
    \label{fig:best_temp-change}
\end{figure}

The lowest temperature applying this method was achieved with the parameter set
\begin{equation}
    \label{eq:2d-temp-change}
    \Delta_1=-405\ \mathrm{GHz}, \quad \Delta_2=-75\ \mathrm{GHz}, \quad \Omega_1=600\ \mathrm{GHz}, \quad \Omega_2=700\ \mathrm{GHz}.
\end{equation}
The minimal temperature was $0.22\ \mathrm{K}$. The time evolution of relevant quantities are shown in Figure~\ref{fig:best_temp-change}. The positive effects of the switching is clear. On average after $0.75/\gamma_a$ time the atoms got trapped $5,000\lambda_1\approx7,5\ \mathrm{mm}$ away from the origin. It is well visible that after this time most of the atoms got trapped in the $x$-direction too, their momentum oscillates around $0$. What stands out in this figure is the rapid decrease of the temperature of the atomic ensemble and under just one lifetime it reaches $1\ \mathrm{K}$. A decent amount of atoms were also kept non-annihilated. $40\%$ of them remained in atomic form after two lifetimes.

\begin{figure}
    \centering
    \includegraphics[width=0.96\linewidth]{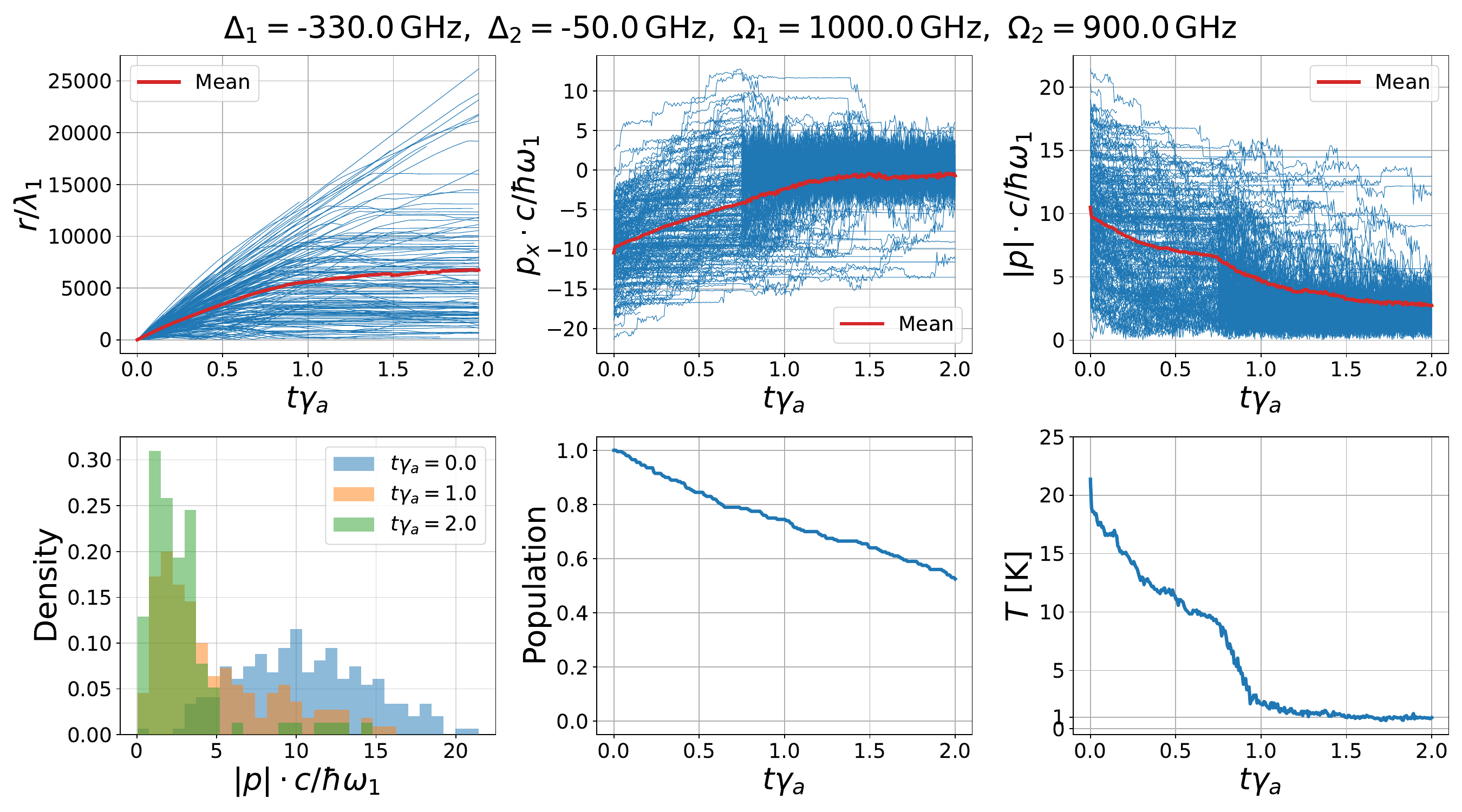}
    \caption{Same as in Figure~\ref{fig:2d-temp}, with laser parameters~\eqref{eq:2d-pop-change} yielding the highest number of non-annihilated atoms at the end of the simulation ($52.5\%$) for the two-dimensional case, when the running wave electric field was changed to a standing one after time $t_r$.}
    \label{fig:best_pop-change}
\end{figure}

The best parameters for keeping the Ps atoms non-annihilated were
\begin{equation}
    \label{eq:2d-pop-change}
    \Delta_1=-330\ \mathrm{GHz}, \quad \Delta_2=-50\ \mathrm{GHz}, \quad \Omega_1=1000\ \mathrm{GHz}, \quad \Omega_2=900\ \mathrm{GHz}.
\end{equation}
This way $52,5\%$ of them were preserved in this form. Figure~\ref{fig:best_pop-change} shows that trapping was achieved in this case too. Temperature did not decrease as quickly as in the previous example, but it still dropped to $1\ \mathrm{K}$, while more than half of the original atoms were kept non-annihilated.

\begin{figure}
    \centering
    \includegraphics[width=0.96\linewidth]{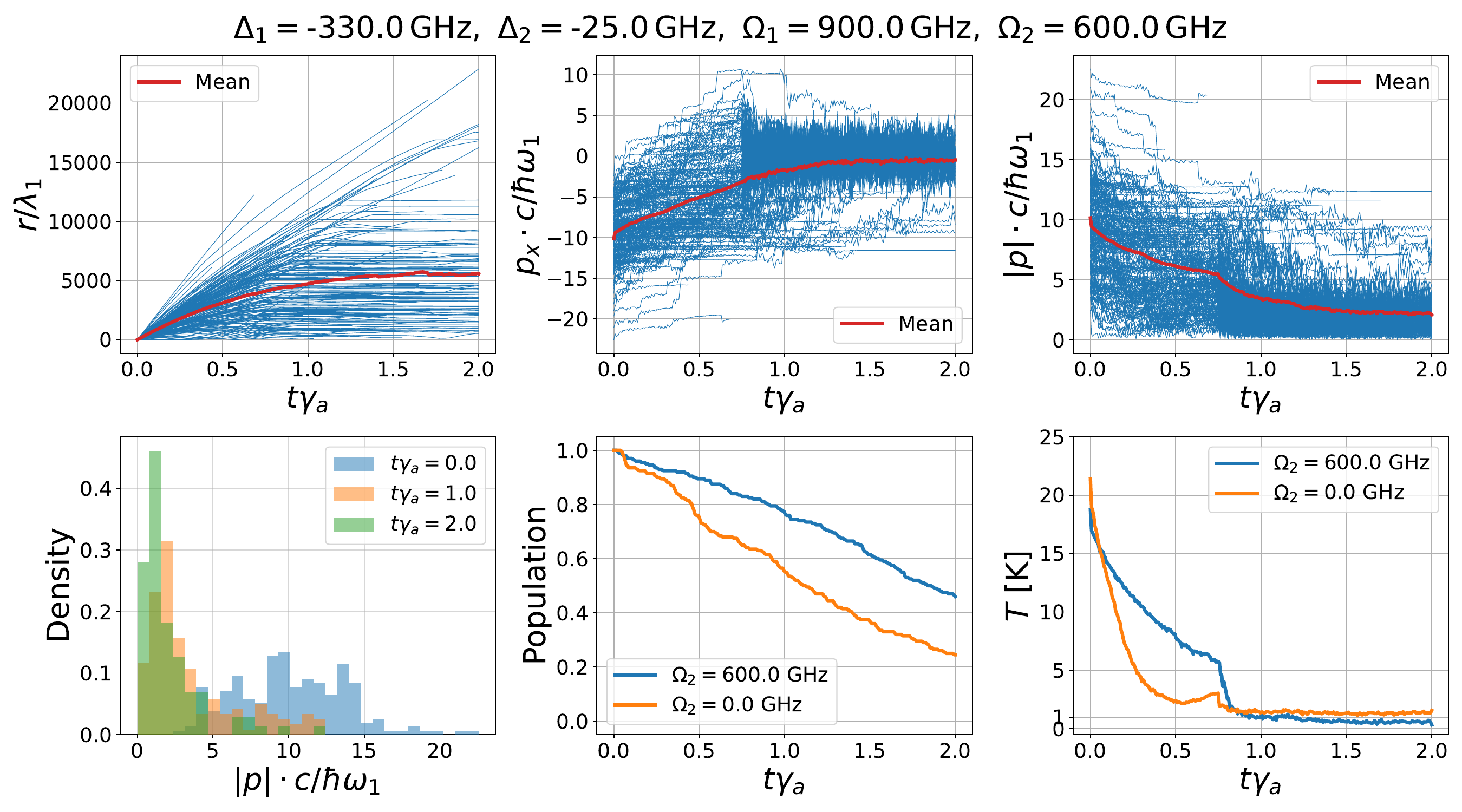}
    \caption{Same as in Figure~\ref{fig:2d-temp}, with laser parameters~\eqref{eq:2d-mix-change}. This resulted in the highest number of non-annihilated atoms when temperature dropped below $1\ \mathrm{K}$ ($79.5\%$) for the two-dimensional case, when the running wave electric field was changed to a standing one after time $t_r$. For the graphs of population and temperature the results of an example run without the $y$-directional laser field are also shown. It is clear from them that this laser beam helps greatly in keeping the atoms non-annihilated and cooling them to lower temperatures.}
    \label{fig:best_mixed-change}
\end{figure}

We also looked for the highest population at $1\ \mathrm{K}$. The optimal laser parameters were
\begin{equation}
    \label{eq:2d-mix-change}
    \Delta_1=-330\ \mathrm{GHz}, \quad \Delta_2=-25\ \mathrm{GHz}, \quad \Omega_1=900\ \mathrm{GHz}, \quad \Omega_2=600\ \mathrm{GHz}.
\end{equation}
In this case $79.5\%$ of the atoms were non-annihilated when the ensemble's temperature reached $1\ \mathrm{K}$. The slowing and trapping processes are clearly visible from the graphs in Figure~\ref{fig:best_mixed-change}. It is also notable in this example that almost half of the atoms ($46\%$) remained in atomic form at the end of the process.

To prove the advantage of the laser beam in the $y$-direction we ran additional simulations setting its strength to $\Omega_2=0$. The other parameters were left the same as described in~\eqref{eq:2d-mix-change}. We found that in this case atoms annihilate faster and temperature does not drop below $1\ \mathrm{K}$. These finding are visible also in Figure~\ref{fig:best_mixed-change} where we plotted the non-annihilated atoms' population and the ensemble's temperature for this case with orange lines. These results indicate the strong positive effect of the laser beam in keeping the atoms non-annihilated and at low temperature. What else stands out in the figure is the time evolution of the temperature. In this case it decreases at a higher pace at the beginning than with the second laser turned on. However the temperature finalises at a higher value.

Moreover, we investigated the cases of applying weaker electric fields too. Interestingly here we found that the laser along the $x$-direction does not require so large intensities. This of course means slower cooling, but the other laser's Rabi frequency can be increased further while having low laser power. This is because the dipole moment of the $\ket{2}\leftrightarrow\ket{3}$ transition is much larger. With these parameter ranges one can possibly keep the atoms non-annihilated for longer, while the ensemble cools down slowly to around $1\ \mathrm{K}$ temperature.

\begin{figure}
    \centering
    \includegraphics[width=\linewidth]{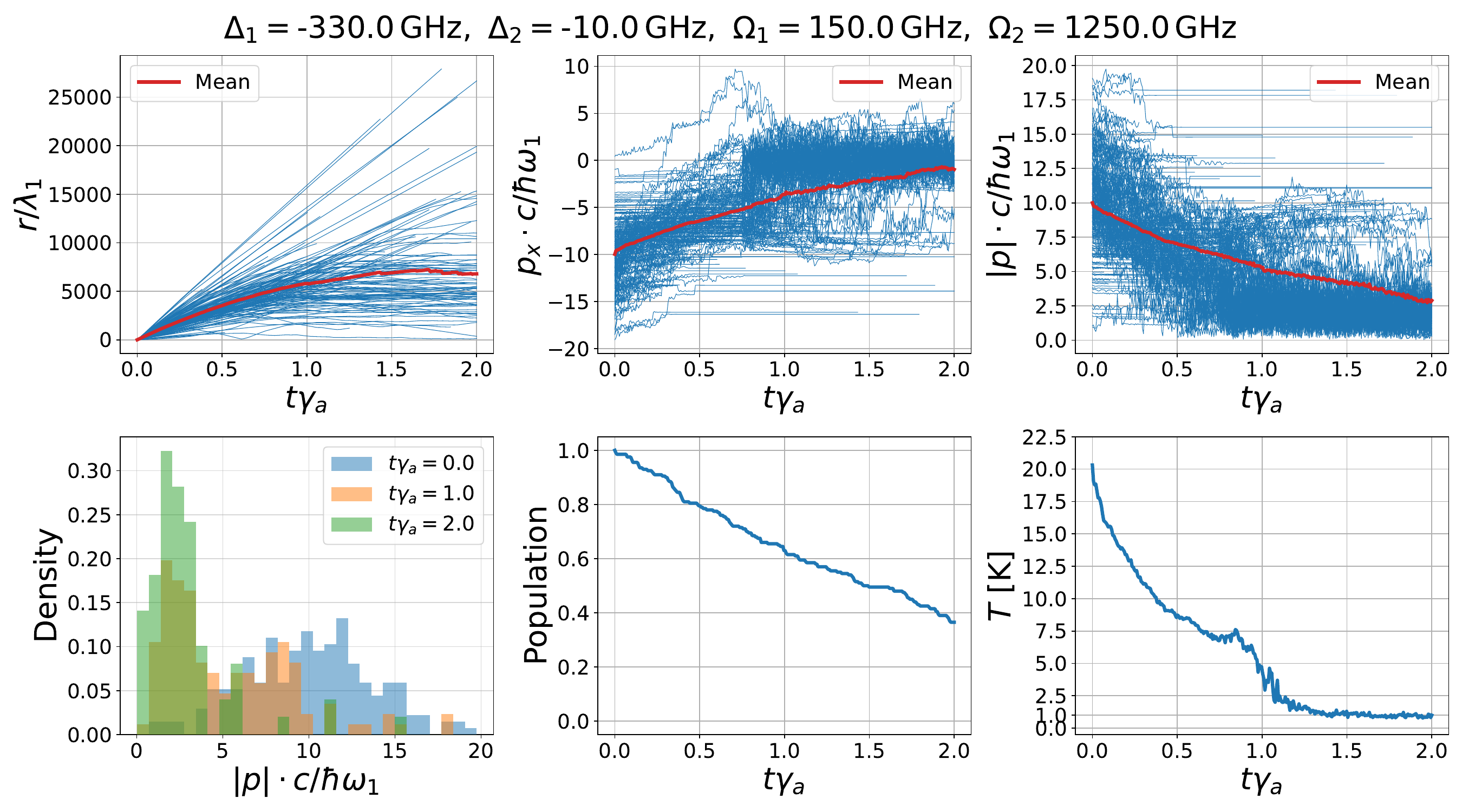}
    \caption{Time evolution of physical parameters described in the caption of Figure~\ref{fig:2d-temp}, with laser parameters~\eqref{eq:2d-low}. This resulted in the highest number of non-annihilated atoms when temperature dropped below $1\ \mathrm{K}$ ($54\%$) for the two-dimensional case, when the running wave electric field's strength was much lower. After time $t_r$ this was changed to a standing wave laser field similarly to the previous cases.}
    \label{fig:best_mixed-low}
\end{figure}

To examine this case we ran another parameter scan in the intervals
\begin{itemize}

    \item $\Delta_1\in\left[-430\ \mathrm{GHz}; -330\ \mathrm{GHz}\right]$,
    \item $\Delta_2\in\left[-50\ \mathrm{GHz}; -10\ \mathrm{GHz}\right]$,
    \item $\Omega_1\in\left[50\ \mathrm{GHz}; 250\ \mathrm{GHz}\right]$, and
    \item $\Omega_2\in\left[1000\ \mathrm{GHz}; 2000\ \mathrm{GHz}\right]$
    
\end{itemize}
in five-five equally distributed points for each. The most promising example was with the following parameters
\begin{equation}
    \label{eq:2d-low}
    \Delta_1=-330\ \mathrm{GHz}, \quad \Delta_2=-10\ \mathrm{GHz}, \quad \Omega_1=150\ \mathrm{GHz}, \quad \Omega_2=1250\ \mathrm{GHz}.
\end{equation}
The relevant physical variables' time evolution is visible in Figure~\ref{fig:best_mixed-low}. It is not surprising that the atoms' deceleration is slower, but after time $t_r$ trapping of the majority of them is clear. However temperature still dropped below $1\ \mathrm{K}$ in under two lifetimes. At the lowest $0.77\ \mathrm{K}$ was reached. With $54\%$ of the atoms non-annihilated the ensemble's temperature decreased to $1\ \mathrm{K}$ and $36.5\%$ remained in the atomic form until the simulation's end.

These results show that the introduction of the $2^3\mathrm{P}$--$3^3\mathrm{D}$ transition can lead to significantly enhanced Ps laser cooling properties even at low laser powers. The intensities required to reach temperature of $1\ \mathrm{K}$ for the last example do not exceed $2.25\cdot10^5\ \mathrm{W}/\mathrm{cm}^2$. Note that we only used a generic geometry here and some quantitative improvements should definitely be possible by a better optimized geometry with several lasers on the upper transition for more lossless cooling. 

\section{Discussion}
\label{discussion}

This study set out to find effective ways to decelerate, cool, and trap ortho-positronium atoms with laser beams. We conducted numerical simulations to find proper parameters for two different models and cooling schemes. The first one involved a rich spectrum of quantum states and one-dimensional motion. We showed that in this case the proper application of a running-, and a standing wave electric field - generated by a laser beam - can result in reaching temperatures well below $1\ \mathrm{K}$. We also found that a large portion of atoms can be kept in a non-annihilated state for reasonably long times. With sufficiently strong electric field we could reach $1\ \mathrm{K}$ temperature with $69\%$ of the initial atoms.

In the second case we examined a simpler $3$-level model for the atoms and two-dimensional motion. Here we wanted to harness the presence of the $3^3\mathrm{D}$ state in order to keep the atoms non-annihilated for longer. At first we considered a running-, and a standing wave electric field as laser scheme. We found that the strong running wave used for slowing re-accelerates the atoms after some time and effective cooling was not achieved.

Moreover we investigated the effects of changing the running wave electric field to a standing one during the process. Our study shows that this helps greatly in the trapping and further cooling of the ensemble of atoms. In the most effective example the temperature dropped below $1\ \mathrm{K}$ while almost $80\%$ of the atoms have not annihilated.

Our research did not cover techniques for keeping the Ps atoms in the non-annihilated states to conduct measurements on them. We point out that some methods have already been proposed for that with the application of dark-states~\cite{gangl2001} or collective radiation~\cite{cui2012}.

These results suggest the possibility of fast Positronium laser slowing, cooling and optical trapping once sufficient UV laser power in the order of $10^7\ \mathrm{W}/\mathrm{cm}^2$ in an approximately $100\ \mathrm{ns}$ pulse is available. We showed that with experimentally feasible parameters and simple laser setups one can achieve sub-Kelvin temperatures for ensembles of reasonable number of atoms starting from pulsed hot source. Our study also found that satisfactory cooling can be achieved with laser intensities in the order of $10^5\ \mathrm{W}/\mathrm{cm}^2$, but more atoms get annihilated in these cases. The generated cold ensemble should allow for a significant improvement of precision spectroscopic QED tests reaching the accuracy of current theoretical modelling. Future prospects of cold Ps based gamma ray amplification and the realisation of Bose--Einstein condensation will, however, need further improved cooling schemes as coherent pulse cooling~\cite{malamant2024coherent}, cavity enhanced or collective cooling. 

\section*{Acknowledgements}

We thank A. Camper, S. Mariazzi and M. Zawada for helpful comments.

Funded by the European Union (QuRIOUS, GA 101227522). Views and opinions expressed are however those of the author(s) only and do not necessarily reflect those of the European Union. Neither the European Union nor the granting authority can be held responsible for them.\\
\includegraphics[width=0.3\textwidth]{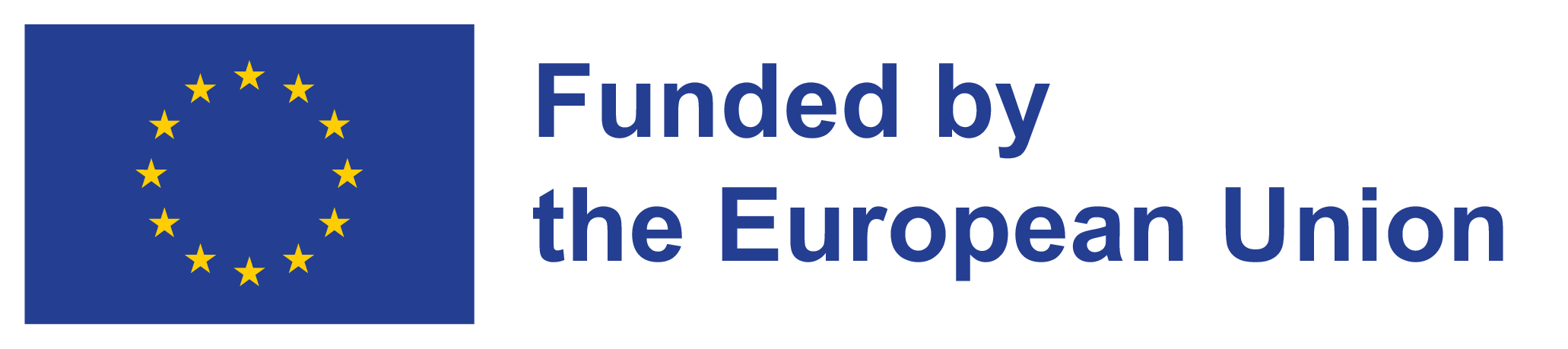}

\newpage
\nocite{*}

\printbibliography

\end{document}